\documentclass[12pt,preprint]{aastex}

\begin{document}

\title{Galactic Globular Clusters with Luminous X-Ray Binaries}

\author{Joel N. Bregman, Jimmy A. Irwin, Patrick Seitzer, and Matt Flores}

\affil{Department of Astronomy, University of Michigan, Ann Arbor, MI 48109}
\email{jbregman@umich.edu}

\begin{abstract}
Luminous X-ray binaries ($>$10$^{{\rm 34}}$ erg s$^{{\rm -}{1}}$, LMXBs) 
have a neutron star or black hole primary, and in globular clusters, most 
of these close binaries are expected to be have evolved from wider binaries 
through dynamical interactions with other stars.  
We sought to find a predictor of this formation rate that is representative of 
the initial properties of globular clusters rather than of the highly evolved
core quantities.
Models indicate the half-light quantities best reflect the initial conditions,
so we examine whether the associated dynamical interaction rate, proportional to 
{\it L$^{{\rm 1.5}}$\/}/$r_{h}^{2.5}$, is useful in understanding the
presence of luminous LMXBs in the Galactic globular cluster system.
We find that while LMXB clusters 
with large values of {\it L$^{{\rm 1.5}}$\/}/$r_{h}^{2.5}$ preferentially 
host LMXBs, the systems must also have half-mass relaxation times below 
t$_{{\rm h,relax}}$ $\sim$10$^{{\rm 9}}$ yr.  This relaxation time effect 
probably occurs because several relaxation times are required to modify 
binary separations, a timescale that must be shorter than cluster ages.  
The frequency of finding an LMXB cluster is enhanced if the cluster is 
metal-rich and if it is close to the bulge region.  The dependence upon 
metallicity is most likely due either to differing initial mass functions 
at the high mass end, or because bulge systems evolve more rapidly from
tidal interactions with the bulge.  This approach can be used to investigate
globular cluster systems in external galaxies, where core properties are
unresolved.
\end{abstract}


\section{Introduction}

X-ray observations can be effective in identifying neutron stars or black holes, since when
they are close to a binary companion, Roche lobe overflow from this secondary leads to a
radiating accretion disk.  In the Milky Way, most of the luminous X-ray binaries lie in the
disk and were born to become X-ray binaries.  That is, the initial separation of the two stars was
sufficiently close that through the course of stellar evolution, it became a Roche Lobe overflow
binary system with a compact primary.  When the initial separation is too great, a
Roche Lobe overflow X-ray binary will not
occur unless gravitational encounters between other stars can redistribute angular momentum,
reducing the separation.  This mechanism is important only in dense stellar systems, such as
globular clusters, where gravitational interactions can occur in less than a Hubble time.

Globular clusters are hundreds of times more likely to host a luminous low mass X-ray
binary (LMXB) than an
equivalent number of field stars in either the disk or the halo \citep{hut92}.  One of the
challenges has been to understand how LMXBs are produced in globular clusters, for which there
has been a great deal of theoretical work.  The central region of a globular has the shortest
relaxation timescale, so it is this region that contracts most rapidly, ultimately undergoing core
collapse, leading to very high central stellar densities.  When it was erroneously believed that
globular clusters had no binary stars, it was posited that hard binaries would be formed in the
central core collapse region.  However, in the past decade, it has become clear that globular
clusters are born with a significant number of binaries (e.g., \citealt{rube97}), which
changed the evolutionary scenario \citep{freg03}.

Binary stars with orbital velocities less than the velocity dispersion of the globular cluster
(soft binaries) are destroyed rather quickly, but binaries with larger orbital velocities (hard
binaries) are destroyed more slowly and their destruction slows the process of core collapse.  The
interaction between stars occurs on the relaxation timescale, which at the half-mass radius, is
about 10$^{{\rm 8}}$-10$^{{\rm 9}}$ yr.  The steady destruction of the binaries occurs over about 5-30 relaxation times, about the time for core collapse.  
The interaction between binaries, and between a binary and a
single star or a compact object can lead to close (harder) binaries where mass transfer becomes
possible (e.g., \citealt{hut92,freg03}).  Binaries that are high-luminosity X-ray
emitters ($>$10$^{{\rm 35}}$ erg/s) have neutron star primaries, so one binary component was
initially a massive star.  Most of the dim sources ($<$ 10$^{{\rm 32}}$ erg/s) are CVs (white dwarf primaries),
although some may be neutron star transients in their low state, especially 
in the 10$^{{\rm 32}}$ - 10$^{{\rm 34}}$ erg/s
range \citep{verb01}.  So the nature of the binaries that one finds depends on the evolutionary
state of the star cluster as well as the mass of the secondary stars that evolve off the Main
Sequence.  That is, age plays a role in both the dynamical state of the globular cluster and the
stellar evolutionary state of the stars that could become mass-transfer binaries.

All-sky surveys, as well as pointed observations with several instruments have discovered
luminous point sources ($>$10$^{{\rm 35}}$ erg/s) in 12 Galactic globular clusters 
(e.g., \citealt{hut92}); henceforth, we only consider these luminous LMXBs.  
The radial distribution of these LMXBs within
the clusters are consistent with being neutron stars rather than massive black 
holes (e.g., \citealt{grind93,grind01}).
It has also been pointed out that these often occur in clusters designated as core
collapse systems, supporting the general picture that enhanced stellar interactions lead to LMXBs. 
Also, it was noted that the metal-rich systems are more likely to host an LMXB
\citep{grind93}, an effect that is
seen in globular cluster systems in early-type galaxies \citep{kund02}.

An important advantage of analyzing LMXBs in Galactic globular clusters is that there is a
vast database available for the globular cluster system \citep{harris96}, including structural
parameters, metallicities, velocities, and often, ages \citep{sala02}.  With the
availability of this extensive data set, we examine the LMXB host systems in detail, where we test
basic model predictions and quantify various correlations in the data.

\section{Model Expectations and Sample Selection}

Our goal is to estimate a quantity representative of the rate of 
stellar interactions that lead to LMXBs over the lifetime of the
globular cluster.  This quantity can be compared to the properties of
globular clusters with luminous LMXBs to determine if it is a useful
predictor.  As is well-known, the relevant quantities are the density
of single stars, the density of binary stars, and the velocity dispersion
of the globular cluster.  Equivalently, a stellar density can be determined
from the globular cluster luminosity and half-light radius.
In principle, one would like to know these quantities at the time that 
the globular clusters were formed, rather than today, as these can evolve 
with time.  In practice, models show that the half-mass radius of the 
majority of the single stars does not evolve strongly with time until
the final destruction of the cluster (see Fig. 8 in \citealt{freg03}), 
unlike the core radius, which can change considerably and is a poor 
indicator of the initial properties of the cluster.  Also,
although the number of stars decreases with time as the cluster evolves, 
the number of stars changes by only about 30\% at the time of the first 
core collapse.  These evolutionary variations are smaller than the range 
of values between globular clusters in the Milky Way:  more than two
orders of magnitude for {\it L;\/} and a factor of 30 for {\it r$_{{\rm h}}$\/}.  
Therefore, {\it r$_{{\rm h}}$\/} and {\it L\/} should be extremely
useful indicators of the initial size and mass of a globular cluster.  
There may be some globular clusters presently in the final stages of destruction
(e.g., NGC 6712; \citealt{demar99,andr01,palt01}), in which case there are no 
good measures of their original properties, but such clusters should 
comprise only a small fraction of the total cluster population.  
The above relationship for the interaction rate has the advantage that
it can be applied to globular clusters in nearby galaxies as well as 
in the Milky Way, since the half-light radius can be fit for galaxies
as distant as the Virgo cluster (using the {\it Hubble Space Telescope};
\citealt{jord05}).

If binaries are transformed into close mass-transfer systems by interactions 
with passing stars, the formation rate is proportional to the gravitational 
interaction rate, which is proportional to 
{\it v$\,$n$_b$$\,$n\/}$\sigma_b$ ($\sigma_b$ is the interaction cross section, 
{\it v\/} is the interaction velocity, {\it n\/} is the density of single
stars, and {\it n$_b$\/} is the density of binary stars).
Most of the interactions that produce the close binaries are between
single stars and binary stars and where the differential velocity is less
than the initial binary orbital velocity \citep{hut83,freg04}, 
in which case $\sigma_b$ $\propto$ {\it v$^{-2}$}.
If, at early time, the binary fraction {\it f\/} was similar between globular
clusters (an assumption that we examine again later), then 
$n_{b}=\frac{f}{1-f}\,n$, and we can express the interaction rate
with a $n^2$ term and an unknown binary fraction.
One can integrate the interaction rate over the volume, and for a 
constant mass-to-light ratio, the total rate is proportional to 
{\it L$^{\rm 2}$ v$^{-1}$ ${r_{\rm h}}^{-3}$\/}$\frac{f}{1-f}$ where 
{\it L\/} is the optical luminosity of the globular cluster 
(proportional to the number of stars) and {\it r$_{{\rm h}}$\/} 
is the half-mass radius of the globular cluster.
We can identify the stellar velocity {\it v} with the velocity dispersion of
the cluster $\sigma_*$, which is related to {\it r$_{{\rm h}}$\/} and {\it L\/}
through the fundamental plane relationships \citep{djor95,mcla00},
$r_{h}^{-0.7}\propto \sigma _{\ast }^{1.45\pm 0.2}L^{-0.85\pm 0.1}$,
which, to within the uncertainties, is the same as the scaling law expected
from the Virial Theorem for a constant mass-to-light ratio, 
{\it r$_{{\rm h}}$\/} $\propto$ {\it L\/} $\sigma{_*}^{-2}$.
Using the scaling from the Virial Theorem, we find that the total 
interaction rate is proportional to {\it L$^{{\rm 1.5}}$\/} $r_{h}^{-2.5} \frac{f}{1-f}$.

The optical catalog of \citep{harris96} contains 147 clusters, but several do not have some
of the structural properties that we need (or they are poorly defined), such as {\it r$_{{\rm h}}$\/} or the core radius,
{\it r$_{{\rm c}}$\/}.  Removing these objects reduces the number of clusters to 141.  The 12 globular clusters with
luminous LMXB have been discussed by (e.g., \citealt{hut92,verb01}, and references therein).  
They are not constant brightness sources, and the time history of their luminosity is not
extensive, so we do not include the X-ray luminosity in our statistical analysis.  
However, many of these sources are transients and visible in X-rays only a
fraction of the time, which needs to be considered in the analysis.
Luminous X-ray sources ($>$10$^{{\rm 36}}$ erg/s) were detectable at 
considerable distances with the {\it ROSAT\/} all-sky survey (to
90 kpc for the survey limit; \citealt{voge00}), so most sources would have been detected.  The
detection rate of LMXBs is not greatly diminished by gas in the Galactic disk, since most of the
luminosity of LMXBs emerges above 1 keV, where absorption is not significant for most sight lines.

\section{Analysis of Globular Cluster Properties}

Many of the optical structural properties of globular clusters have been discussed in the
literature (e.g., \citealt{ash98}), so what we add is a quantitative analysis of whether the globular clusters that host
LMXBs are distinguished in some way.  One of the expectations is that as clusters evolve, their tidal
radius is determined by environmental factors, such as whether the cluster has passed through the
disk or passed by the inner bulge of the Galaxy.  The tidal radius should be largely decoupled
from the inner part of the cluster, where most of the LMXBs would be formed, so we would not
expect a good relationship between the presence of LMXBs and {\it r$_{{\rm t}}$\/} (or {\it r$_{{\rm t}}$\/} normalized by {\it r$_{{\rm h}}$\/}, the ratio
{\it r$_{{\rm t}}$\/}/{\it r$_{{\rm h}}$\/}).  This expectation is confirmed in that the presence of an LMXB is randomly distributed as a
function of {\it r$_{{\rm t}}$\/}/{\it r$_{{\rm h}}$\/}, as seen in the histogram (Figure 1).  Henceforth, we concentrate on
quantities other than {\it r$_{{\rm t}}$\/} in this investigation.

For unevolved clusters, we would expect a linear correlation between {\it r$_{{\rm h}}$\/} and {\it r$_{{\rm c}}$\/}, but as the
clusters evolve dynamically, {\it r$_{{\rm c}}$\/} is predicted to change much more than {\it r$_{{\rm h}}$\/}.  Therefore, the smallest
clusters (as measured by {\it r$_{{\rm h}}$\/}) often have the shortest relaxation times, so the departure from the
original relationship between {\it r$_{{\rm h}}$\/} and {\it r$_{{\rm c}}$\/} should be greatest, and this is evident in Figure 2. 
At the larger values of {\it r$_{{\rm h}}$\/}, there is a narrow and well-defined relationship with {\it r$_{{\rm c}}$\/}, with a dispersion
of less than a factor of two.  However, at the smaller values of {\it r$_{{\rm h}}$\/}, there is a very large range in {\it r$_{{\rm c}}$\/},
with a range of about an order of magnitude and extending only toward small values of {\it r$_{{\rm c}}$\/}.  
Only one of the LMXBs occurs in a cluster that lies on the 
unevolved {\it r$_{{\rm c}}$\/} - {\it r$_{{\rm h}}$\/} relationship, with the other
11 having values of {\it r$_{{\rm c}}$\/} that are 3-30 times smaller 
than {\it r$_{{\rm c}}$\/} for the unevolved relationship.
This one outlier is NGC 6712, which as mentioned above, is probably
in the process of being disrupted \citep{palt01}, so its value of
{\it r$_{{\rm h}}$\/} is likely not representative of its value at
formation.

Because {\it r$_{{\rm c}}$\/} can change by very large factors during the evolution of the cluster, it is a poor
indicator of the original state of the cluster, and its value can be uncertain due to the small number
of stars involved in its determination.  The best indicator of the original size of the cluster is {\it r$_{{\rm h}}$\/},
while the best indicator of the total number of stars is M(V).  Neither are perfect indicators, as
they can differ significantly from their original values when the cluster is being tidally torn apart. 
Nevertheless, we can use these two properties to estimate the interaction rate (discussed above),
where we find that clusters hosting LMXBs tend to be small and luminous, preferentially with high
interaction rates, given by {\it L$^{{\rm 1.5}}$\/}/$r_{h}^{2.5}$ (Figure 3;
we have dropped the factor due to binary fraction and will return to that issue).  
A t-test shows that the distribution of interaction rates differs at greater
than the 99\% level between clusters hosting LMXBs and those without luminous LMXBs.
Closer inspection of this figure shows that
the lines of constant interaction rates are not ideal predictors of the presence of an LMXB.  At
constant interaction rate, the smaller clusters appear to be more likely than the larger ones to host
an LMXB, so some modification of the nominal prediction is implied.

Although the core radius is a current property rather 
than an original property of the cluster, we see the strong 
preponderance for clusters with small {\it r$_{{\rm c}}$\/} to host LMXBs, as
has been noted previously (Figure 4;
note that the range in radius is much larger than when using {\it r$_{{\rm h}}$\/}).  
Many of the systems with very small {\it r$_{{\rm c}}$\/} 
have undergone core collapse, or may have undergone core collapse, 
as this can be difficult to identify if there has been a core ``bounce'' 
as most models predict.  Also, for the less luminous systems, 
few stars comprise the core so it can be difficult to obtain 
a reliable value of {\it r$_{{\rm c}}$\/}.  Finally, a number of
systems have power-law density distributions into the smallest observable
radii, so the concept of a core radius becomes less useful \citep{noyo03}.

Another approach to examining these clusters is by using the relaxation time defined at the
half-mass radius.  With the relaxation time {\it t$_{{\rm h,relax}}$\/} $\propto$ {\it L$^{{\rm 0.5}}$\/} $r_{h}^{1.5}$, we can form a stellar interaction rate
for the number of stars in the cluster, {\it N$_{{\rm 0}}$ /t$_{{\rm h,relax}}$\/} $\propto$ {\it L$^{{\rm 0.5}}$\/}/ $r_{h}^{1.5}$, which has a steeper relationship
between {\it L\/} and {\it r$_{{\rm h}}$\/} at a constant rate ({\it L\/} $\propto$ {\it r$_{{\rm h}}$\/}$^{{\rm 3}}$) than does the two-body interaction rate ({\it L\/} $\propto$ {\it r$_{{\rm h}}$\/}$^{{\rm 5/3}}$
at constant rate), due to the inclusion of the binary star cross section in the second rate.  When
inspecting the relationship between {\it t$_{{\rm h,relax}}$\/} and M(V) (Figure 5)  there appears to be a
relaxation time, at about 2$\times$10$^{{\rm 9}}$ yr, above which there is only one LMXB, even though about half of
the clusters have larger relaxation times.  This suggests that at least five relaxation times are
required to form enough close binaries that luminous X-ray sources are seen.

A particularly striking relationship is the success of a cluster hosting an LMXB as a function
of central density (Figure 6), which is a current cluster quantity.  Of the optically luminous
systems (M(V) $<$ -7), 8/9 have the highest central densities of the sample, with 5$<$ log$\rho$ $<$ 5.5. 
For these optically luminous clusters, at these high densities, the number of clusters with LMXBs is
greater than those without LMXBs!  Only one cluster hosting an LMXB has a central density that is
near the average value for clusters, the system NGC 6712, which is probably in the process
of being disrupted (as discussed above).
For the next two fainter clusters ($-$7 $<$ M(V) $<$ -5), the LMXB-bearing systems have
central densities that are well above average for the sample, although not quite as dense as the
more luminous systems.  The very faintest cluster with an LMXB has a density near the sample
average, but there are very few clusters in this region and this cluster may be evolving rapidly
now, undergoing destruction.

There are two other properties, involving metallicity and position, that affect the
likelihood of a cluster hosting an LMXB.  The cluster metallicity distribution is bimodal, with  most
clusters being low metallicity (the modal value of the low metallicity component is log[Fe] = $-$1.6
and we use log[Fe] = -1.1 as the division between high and low metallicity systems; 
\citealt{ash98}), yet most of
the clusters with LMXBs are high-metallicity systems (Figure 7), an effect that has been noted
previously in the Galaxy and in early-type galaxies (e.g., \citealt{grind93,kund02}). 
Also, there is a preponderance for clusters hosting LMXBs to lie within 4 kpc of the center of the
Milky Way (Figure 8).  This may be due to the tidal influence of the bulge causing the
bulge population of clusters to evolve more rapidly.  Alternatively, this bulge cluster population
may simply have formed more compact systems initially.

\section{Discussion and Interpretation}

These results are interpreted within the context of models, which have advanced greatly in
the last decade, but are not yet complete in the sense of having a ``standard model'' to use. 
Consequently, we will interpret our results in terms of the generic features of present models,
relying most heavily on those of \citet{freg03}, and in particular, those with King profiles
(W$_{{\rm 0}}$= 7).  These models include binaries with different amounts of initial binary fractions from 2-20\% and they calculate the time variation of various binary properties, such as their destruction
and their hardness distribution.  These models show that as the cluster evolves dynamically, the binaries are hardened (and destroyed), a process that delays core collapse.  The binary fraction
decreases (but hardens) by about half in 10-20 t$_{{\rm h,relax}}$ times, which we consider a characteristic
time.  This implies that when t$_{{\rm h,relax}}$ $\lesssim$ 10$^{{\rm 9}}$ yr, the cluster has significantly modified the initial binary
population, possibly causing close binaries that form into LMXBs, in general agreement with our
findings.

The initial predictor of close binary formation from the rate of binary collisions should
then be modified by the time needed for a cluster to begin evolving, relative to its age.  The
original rate (proportional to {\it L$^{{\rm 1.5}}$\/}/$r_{h}^{2.5}$ might be modified by a term such as (1-{\it exp(t$_{{\rm age}}$/$\eta$t$_{{\rm h,relax}}$\/})),
where $\eta$ $\approx$ 5-10.

Provided that the cluster does not become disrupted, the number of binaries eventually
decreases to the level at which it can no longer prevent core collapse (about 0.2\% binary
fraction).  The core collapses and oscillates even as r$_{{\rm h}}$ hardly changes (e.g., the 5\% binary fraction
model of Fregeau).  During this process, some binaries still exist and are extremely hard and
therefore likely to become mass transfer systems during the course of stellar evolution.  At these
high densities, direct collisions between neutron stars and red giants can produce LMXBs as well
\citep{ivan05}.

Half of the LMXB clusters are designated as core collapse systems (compare to one-fifth for
the non-LMXB clusters), where the definition of a core collapse object is that it has a power-law
optical surface brightness distribution into the center with no apparent core radius (for objects
labeled as ``possible core collapse'', it is difficult to be certain of the power-law distribution into
the center).  We can take a different evolutionary definition on the basis of the models of \citet{freg03}.
In their models, the ratio r$_{{\rm c}}$/r$_{{\rm h}}$ begins near 0.2-0.3 and slowly decrease to about
0.07, after which a core-collapse event causes r$_{{\rm c}}$/r$_{{\rm h}}$ to fall to 0.01 or less.  This deep collapse is
short-lived and the core reexpands, spending most of its time at r$_{{\rm c}}$/r$_{{\rm h}}$ $\approx$ 0.05-0.1 before a
subsequent recollapse.  The quantitative values of this model cannot apply to all clusters since
about half of the clusters have values of r$_{{\rm c}}$/r$_{{\rm h}}$ $>$ 0.3, so we must take caution in making conclusions
based on these models.  If we consider objects with r$_{{\rm c}}$/r$_{{\rm h}}$ $<$ 0.1 to have undergone or about to
undergo core collapse, then 4/12 (33\%) LMXB-clusters are core collapse candidates, compared to
16/129 (12.4\%) for the non-LMXB clusters.  If this identification with core collapse is correct,
having passed through this stage is helpful for the formation of LMXBs, but not essential.  The
more important influence in forming LMXBs may be the many stellar interactions that occur prior
to core collapse.

The preference of finding LMXB clusters toward the inner part of the Galaxy may reflect the
role of the bulge on the evolution of clusters.  Tidal influences are greater in the inner part of the
Galaxy, and this removes the ``hotter'' stars, effectively cooling the cluster and allowing it to
dynamically evolve more rapidly, which can in turn, can produce more close binaries.

Regarding the role of metallicity, our initial expectation was that these are younger
systems and thereby different in their stellar population than the older systems.  The best uniform
study of cluster ages is by \citet{sala02}, where they produce a uniform set of ages for
55 globular clusters.  While they find a mild dependence for some metal-rich clusters being
younger, they also find that the systems within the Solar Circle have the same ages, regardless of
metallicity.  Their sample contains five of our LMXB clusters (NGC 1851, NGC 6624, NGC 6652,
NGC 6712, and NGC 7078), all of which have ages in the range 9.2-11.7 Gyr ($\pm$1.1 Gyr) that are
typical for the rest of the sample.  Since this study rules out age as the important parameter for
producing LMXBs in Galactic globular clusters, there must be some other mechanism at work, 
such as the initial binary fraction, number of high mass stars that become neutron stars (a
difference in the IMF), or the degree of tidal influence on these systems by the bulge.

The separation by metallicity and location of LMXB clusters may reflect the difference of
two types of globular cluster populations, those that formed with the bulge and those that are halo
objects.  The more metal-rich bulge clusters spend more time close to the inner part of the Galaxy,
hence they experience more tidal forces, leading to more rapid evolution and destruction.  Many if
not most of the halo clusters did not form with the bulge and they occupy a larger volume and
have a larger velocity dispersion.  Although they are found at all radii, including R $<$ 4 kpc, they
spend relatively less time in this region due to their orbits having larger semi-major axis.

A difference in the binary fraction between high and low metallicity systems should be
observable, but ideally, one wants to measure the binary fraction in relatively unevolved clusters,
since significant cluster evolution is expected to destroy binaries but leave the remaining ones with
relatively small separations (i.e., only hard binaries remain).  The current set of data for binary
fractions is not uniform, favors low metallicity systems, and has several evolved systems.  
A cluster listed as a core collapse system, NGC 6397 ([Fe/H] = -1.95) has a low binary
fraction, with an upper limit of about 5\% \citep{cool02}.  This is to be expected
since the large majority of the initial binaries should have been destroyed.  
However, another core collapse system is NGC 6752([Fe/H] = -1.56, r$_{{\rm c}}$/r$_{{\rm h}}$ = 0.07), 
yet it has a much larger binary fraction of 15-38\% in the core region \citep{rube97}. 
A less evolved system is NGC 6121 (M4; [Fe/H] = $-$1.20, the most
metal rich of this group of four clusters), with properties that
make its evolution about average relative to the entire globular cluster sample
(r$_{{\rm c}}$/r$_{{\rm h}}$ = 0.23 and t$_{{\rm h,relax}}$ = 6.6$\times$10$^{{\rm 8}}$ yr) but its binary fraction is only 1-2\% \citep{rich04}.   
The least evolved system for which there has been a binary study 
is NGC 288 ([Fe/H] = -1.24), just to the low metallicity side of
our low/high metallicity dividing line, but with r$_{{\rm c}}$/r$_{{\rm h}}$ = 1.0 and t$_{{\rm h,relax}}$ = 6.5$\times$10$^{{\rm 9}}$ yr.  
In NGC 288, \citet{bel02} find a binary fraction of 10-20\% within r$_{{\rm h}}$.
Although considerably more observations need to be
made, especially of relatively unevolved systems, there is no evidence that the lower metallicity
systems are lacking in binaries (e.g., NGC 288 and NGC 6752), 
suggesting that the high metallicity systems had a shallower IMF
at the high mass end or that they evolve more due to Galactic tidal influences.  We note that it
might be possible to use blue stragglers to gain insight into the binary population (e.g., \citealt{piot03}), but these are binaries that have undergone dynamical encounters and are not good
tracers of the initial binary population.

The relationship between the nature of globular cluster densities and the number of
compact objects has been studied using the more sensitive {\it Chandra\/} data, that permit observers to
study sources fainter than 10$^{{\rm 34}}$ erg s$^{{\rm -}{1}}$ (summary in \citealt{hein03}).  In an analysis of these
sources for a dozen clusters, \citet{pool03} found that the number of X-ray sources per
cluster (L$_{{\rm x}}$ $>$ 4$\times$10$^{{\rm 30}}$ erg s$^{{\rm -}{1}}$) was correlated with the total number of stars in the cluster, but more
tightly correlated with the cluster encounter rate.  Their encounter rate is the volume integral of
the local encounter rate, $\rho$$^{{\rm 2}}$/{\it v\/}, which is different from our rate in that it emphasizes the present day
conditions, particularly in the core, which can dominate the integral. Once many clusters are
observed to low luminosities, this approach certainly will be superior to the study presented here
as the statistical measure of hard binaries is better quantified.

Some of the correlations discussed in our study occur in other galaxies, such as the
correlation between the presence of an LMXB and the metallicity (color is the proxy for metallicity
in most extragalactic studies, aside from Local Group galaxies; e.g., \citealt{kund02}).  Yet other
galaxies will permit the studies of the LMXB rate with age, since some galaxies are younger than
others or have globular clusters known to be young (e.g., M31, LMC).  With Galactic globular
clusters, it might be possible to measure the binary fraction rate for high and low metallicity
systems, testing our conclusion.  Also, the number of globular clusters with LMXBs is a modest 12
systems, and the study of certain elliptical galaxies offers the opportunity of increasing the
statistics by an order of magnitude (e.g., \citealt{ang01}; \citealt{irwin03}),
since $r_h$ can be measured in external galaxies at the distance of the Virgo Cluster \citep{jord05}.
Finally, we look forward to the improvement in models, which in principle should be able to
predict the frequency of LMXBs for globular clusters of various initial properties and evolutionary
states.

We would like to acknowledge support for this work from {\it NASA\/} through grant NAG5-10765.  
Also, we would like to thank Mario Mateo, Renato Dupke, Edward Lloyd-Davies, and an
anonymous referee for their valuable advice.

\begin{figure}
\plotone{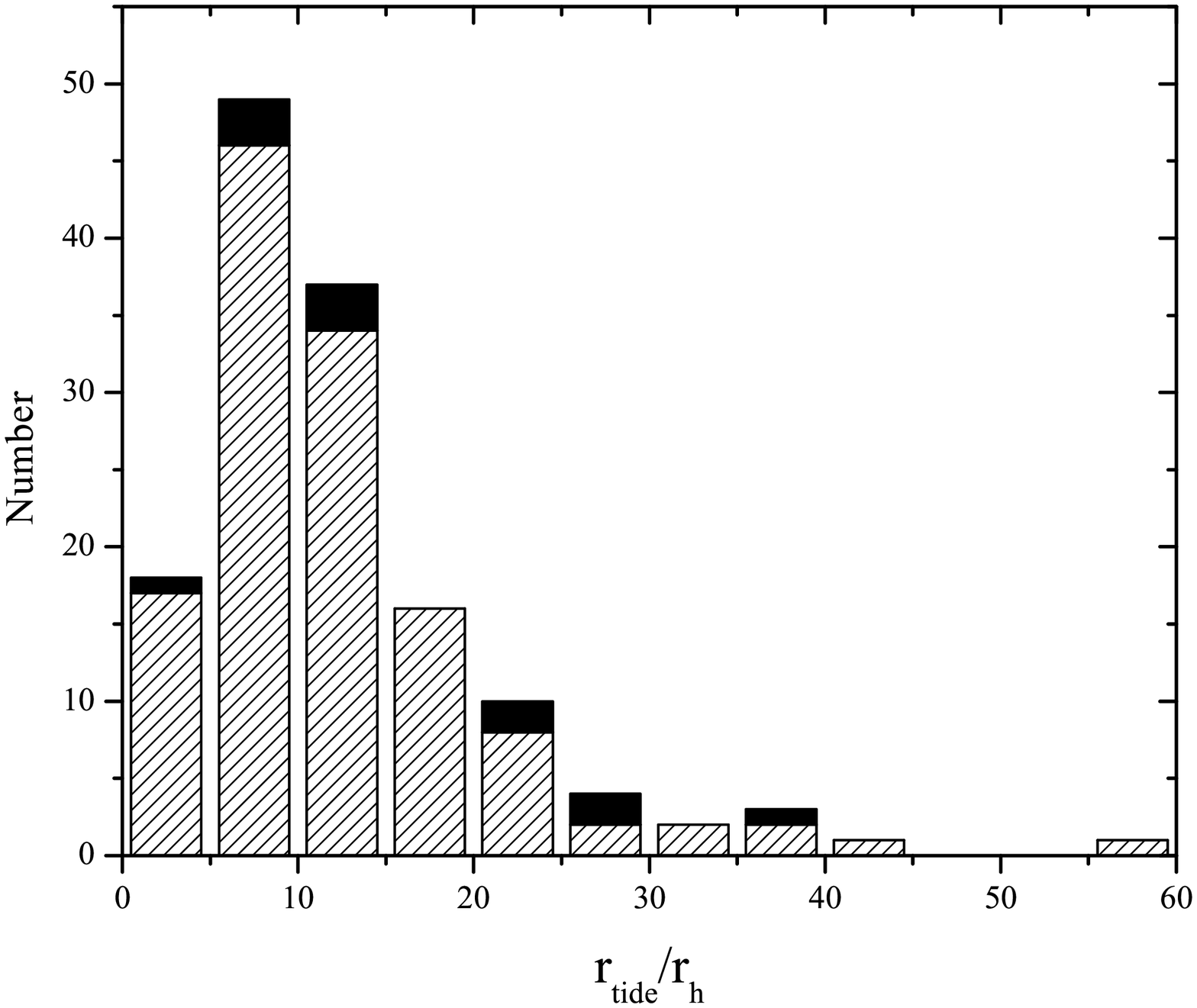}
\caption{The ratio of the tidal radius to half-mass radius for the globular clusters in the
sample.  The hatched regions are clusters without LMXBs while the solid regions are clusters with
LMXBs.}
\end{figure}

\begin{figure}
\plotone{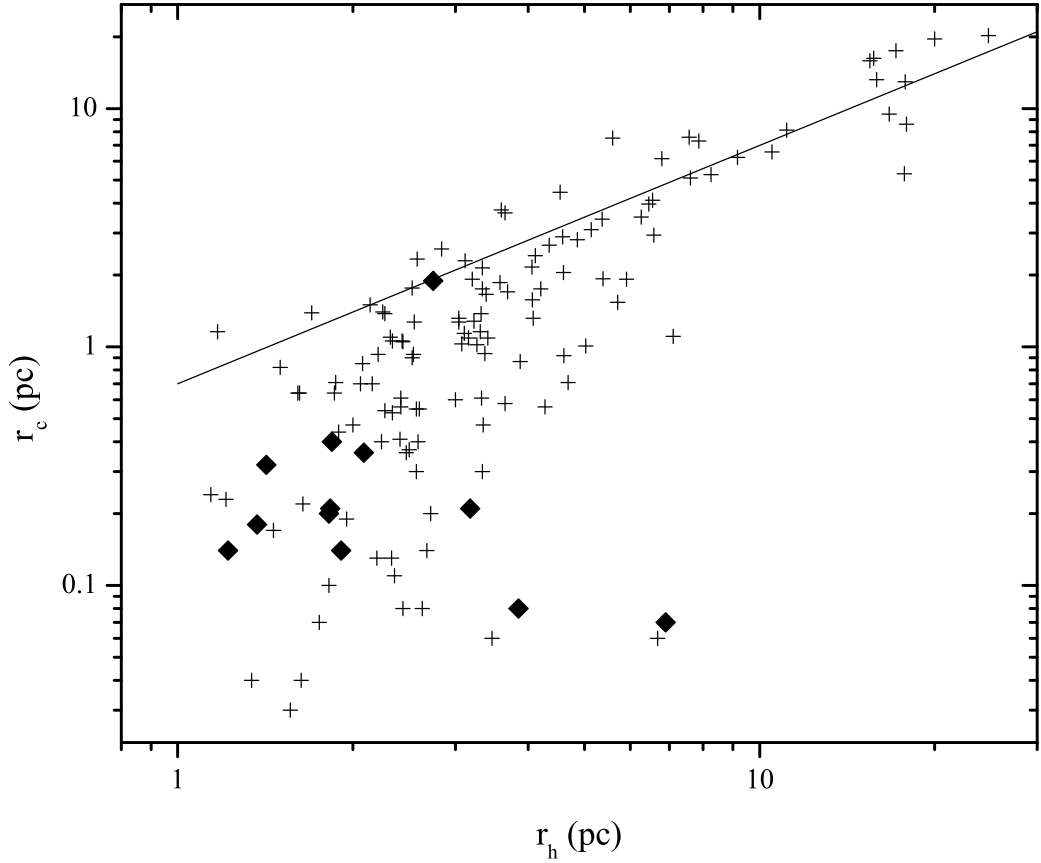}
\caption{The distribution of globular clusters by core radius and half-mass radius,
where the solid diamonds are systems with luminous LMXBs and the crosses are those without
LMXBs.  The line, of slope unity, may indicate the unevolved population of globular clusters, since
r$_{{\rm c}}$ can evolve, becoming much smaller, while r$_{{\rm h}}$ changes relatively little until disruption occurs.}
\end{figure}

\begin{figure}
\plotone{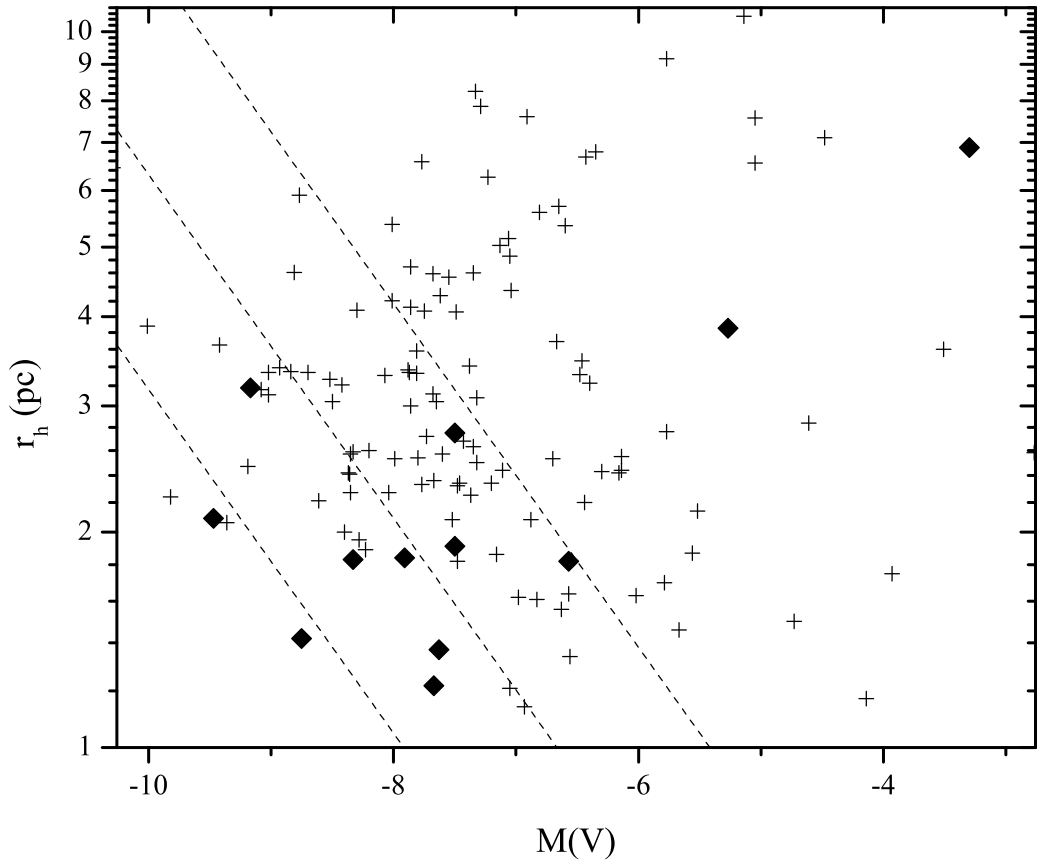}
\caption{The distribution of r$_{{\rm h}}$ and M(V) for the globular cluster sample, where the solid
diamonds are systems with LMXBs.  Lines of constant binary collision rate are shown ({\it L$^{{\rm 1.5}}$\/}/$r_{h}^{2.5}$),
with a difference of 0.5 dex between the lines.  There are two outliers (M(V) $>$ -6), which are
small systems that may be in the process of disruption.  Note that relatively few LMXB clusters
have r$_{{\rm h}}$ $>$ 3 pc, suggesting that the success rate of having 
an LMXB depends more strongly on r$_{{\rm h}}$ than predicted from the collision rate.}
\end{figure}

\begin{figure}
\plotone{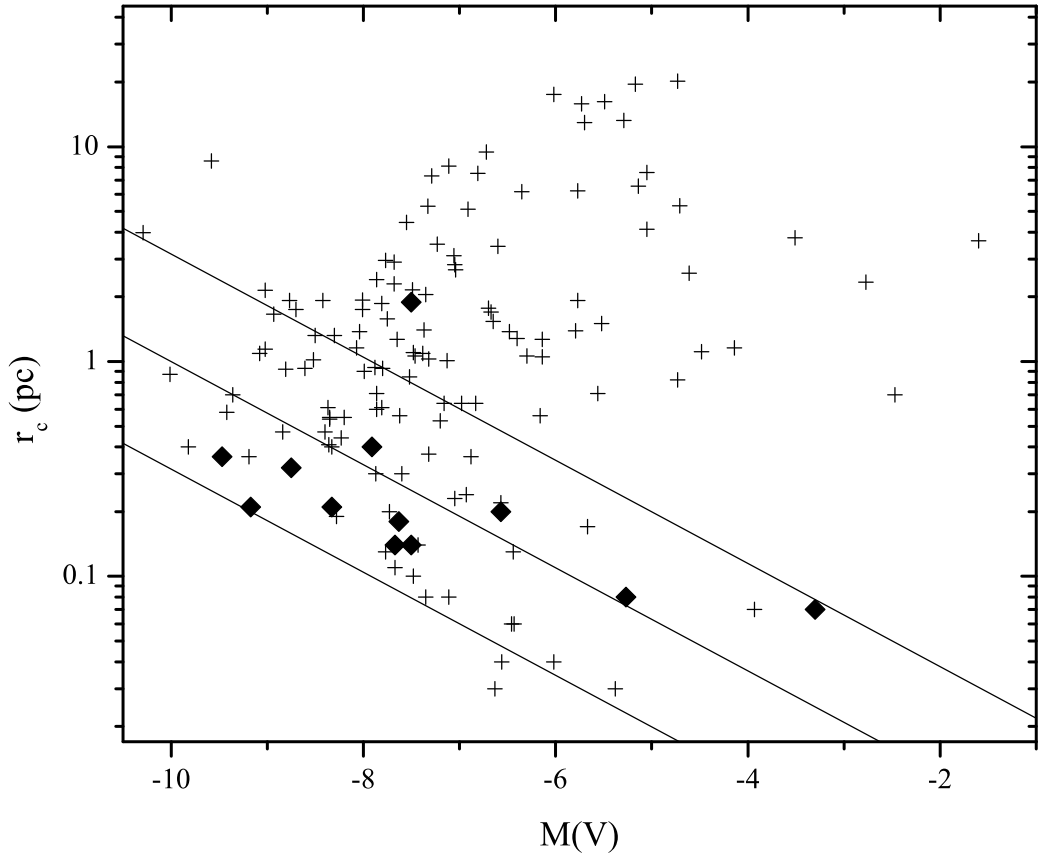}
\caption{The distribution of r$_{{\rm c}}$ and M(V) for the globular cluster 
sample, with the same symbols and lines as in Figure 3.  
Note that the range of r$_{{\rm c}}$ is nearly three orders of
magnitude, compared to one order of magnitude for r$_{{\rm h}}$.}
\end{figure}

\begin{figure}
\plotone{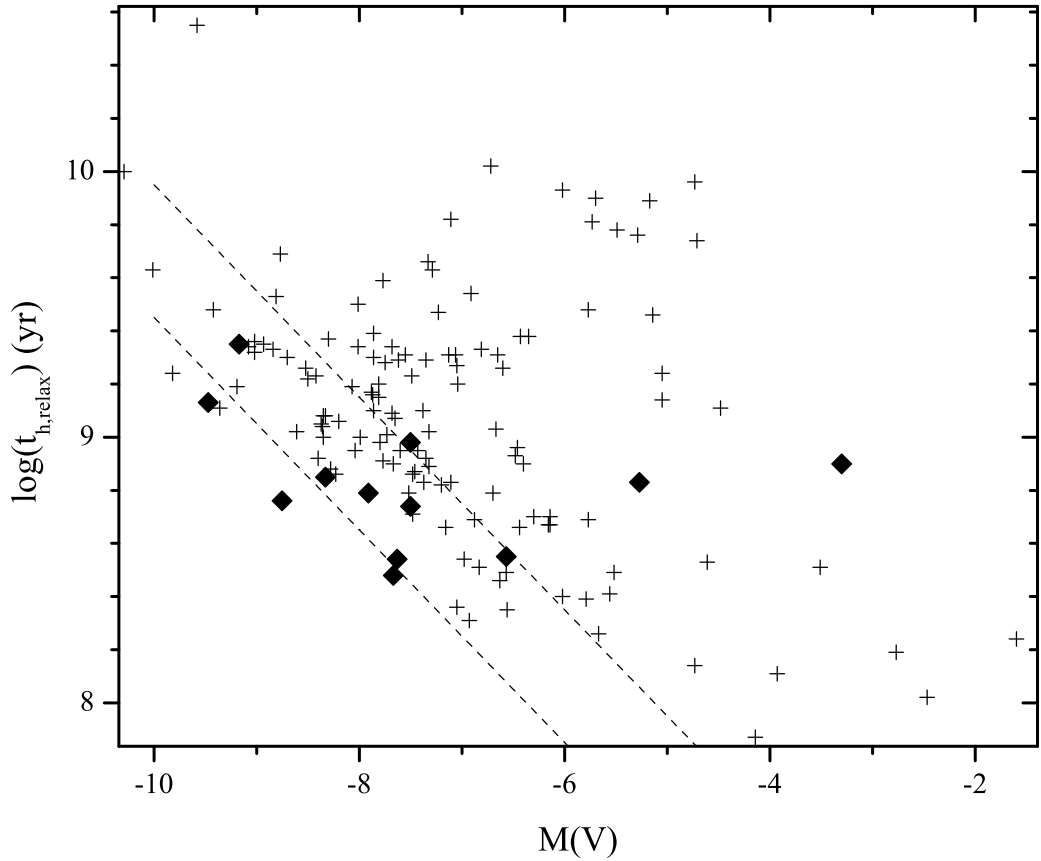}
\caption{The half-mass relaxation time as a function of M(V) for 
the LMXB clusters (diamonds) and non-LMXB clusters (crosses).  
The dotted lines are loci of constant {\it L/t$_{{\rm h,relax}}$\/}, 
with a difference of 0.5 dex between the lines.  Most of the LMXB 
hosts have relatively small values of {\it L/t$_{{\rm h,relax}}$\/}, 
which is similar to, but not the same as the two-body collision rate.}
\end{figure}

\begin{figure}
\plotone{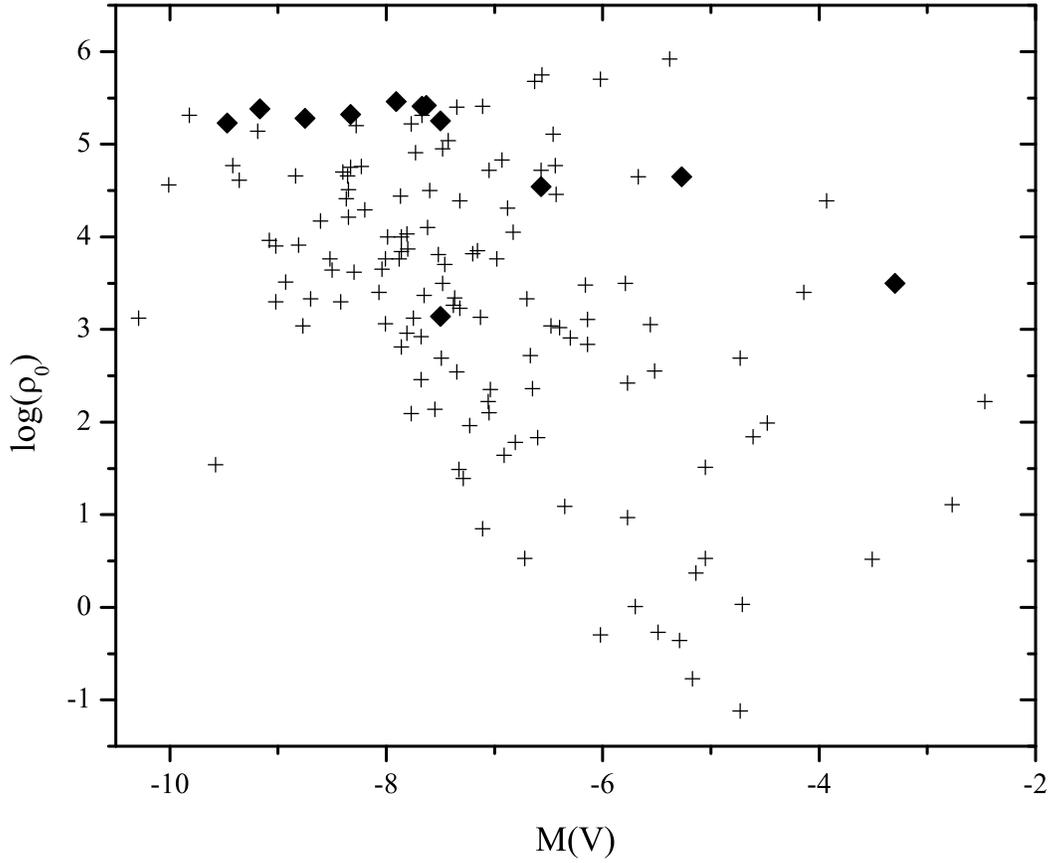}
\caption{The logarithm of central luminosity density (L$_{{\rm \odot}{}}$ pc$^{{\rm -}{3}}$) as a function of absolute
magnitude shows that most clusters that host a luminous LMXB have extreme values o the central
density.}
\end{figure}

\begin{figure}
\plotone{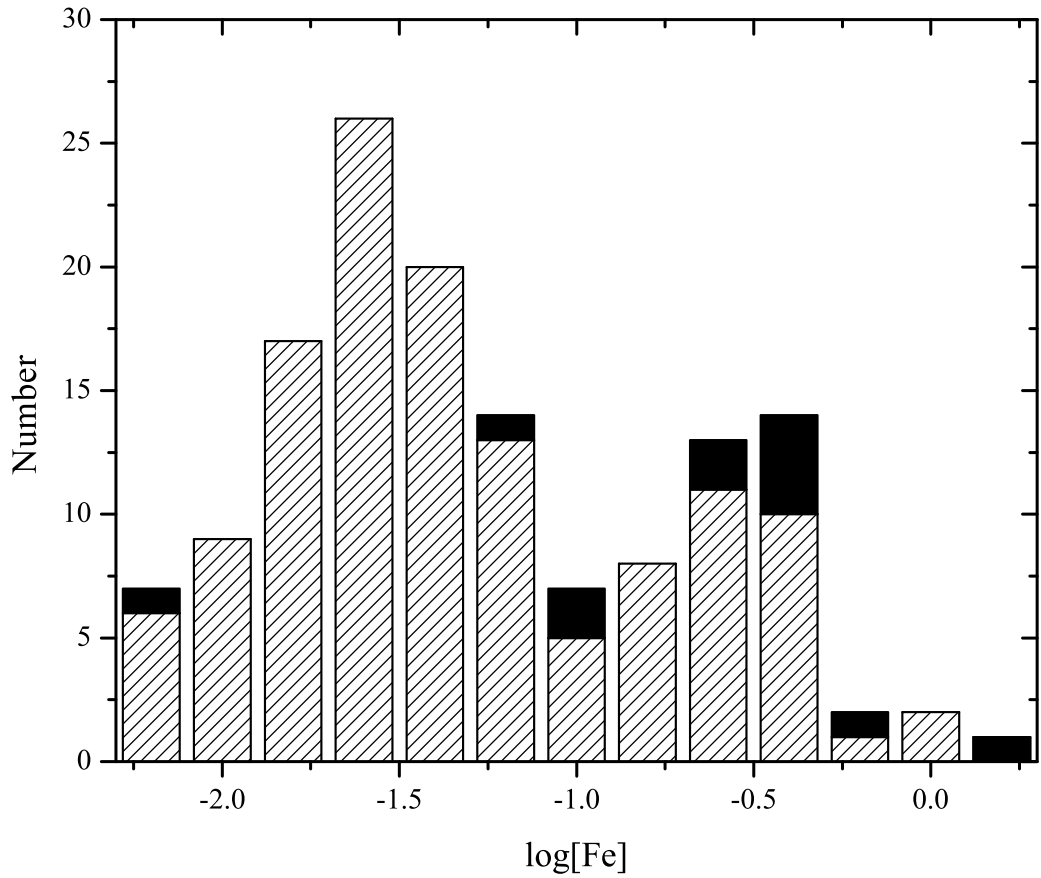}
\caption{The histogram of metallicities (LMXB clusters are solid) with the familiar double-peaked
distribution of high and low metallicity systems shows that most of the LMXB clusters are of
moderate to high metallicity.}
\end{figure}

\begin{figure}
\plotone{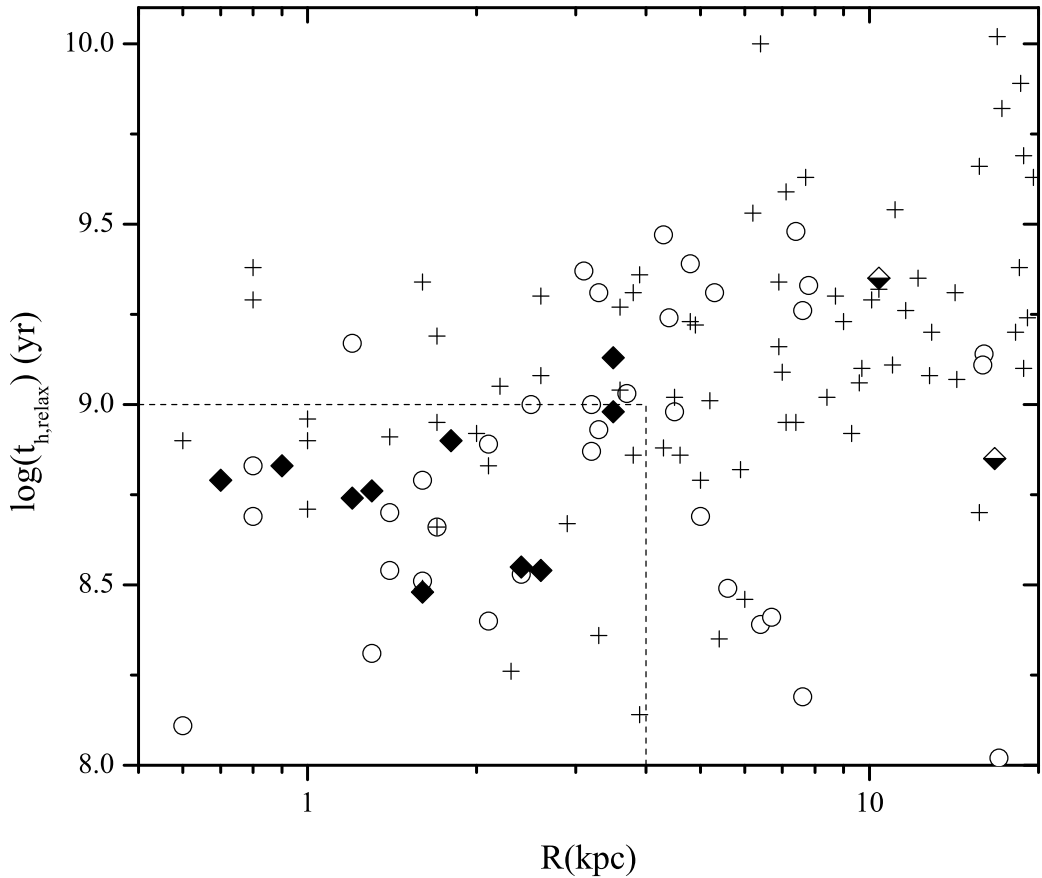}
\caption{The globular cluster sample as a function of t$_{{\rm h,relax}}$ and distance from the center
of the Galaxy (R).  Metal-rich clusters with LMXBs are solid diamonds, metal-poor LMXB clusters
are half-filled diamonds, metal-rich clusters without LMXBs are open circles and metal-poor
clusters without LMXBs are crosses.  Clusters with R $>$ 12 kpc or M(V) $>$ -2 are not shown.  Most
of the LMXB clusters lie in the region R $<$ 4 kpc and t$_{{\rm h,relax}}$ $<$ 10$^{{\rm 9}}$ yr (dashed lines), all of which are
metal-rich clusters.}
\end{figure}


\begin{thebibliography}{} 

\bibitem[Andreuzzi et al.(2001)]{andr01} Andreuzzi, G., De 
Marchi, G., Ferraro, F.~R., Paresce, F., Pulone, L., \& Buonanno, R.\ 2001, 
\aap, 372, 851
\bibitem[Angelini, Loewenstein, \& Mushotzky(2001)]{ang01}Angelini, L., Loewenstein, M., \& Mushotzky, R.F. 2001, \apjl, 557, L35
\bibitem[Ashman \& Zepf(1998)]{ash98} Ashman, K.~M., \& Zepf, 
S.~E.\ 1998, Globular cluster systems; Keith M.~Ashman, Stephen E.~Zepf.~ 
Cambridge, U.~K.~; New York : Cambridge University Press, 1998.
\bibitem[Bellazzini et al.(2002)]{bel02}Bellazzini, M., Fusi Peci, F., Messineo, M., Monaco, L.,
\& Rood, R.T. 2002, \aj, 123, 1509
\bibitem[Cool \& Bolton(2002)]{cool02}Cool, A.M., \& Bolton, A.S. 2002, in Stellar Collisions, Mergers and their Consequences, ASP Conference Proceedings, Vol. 263. Ed. Michael M. Shara, San Francisco: Astronomical Society of the Pacific), 163
\bibitem[de Marchi et al.(1999)]{demar99} de Marchi, G., 
Leibundgut, B., Paresce, F., \& Pulone, L.\ 1999, \aap, 343, L9 
\bibitem[Djorgovski(1995)]{djor95}Djorgovski, S. 1995, \apjl, 438, L29
\bibitem[Fregeau et al.(2004)]{freg04}Fregeau, J.M., Cheung, P., Portegies Zwart, S.F., \& Rasio, F.A. 2004, \mnras, 352, 1
\bibitem[Fregeau et al.(2003)]{freg03}Fregeau, J.M., G\"{u}rkan, M.A., Joshi, K.J., \& Rasio, F.A. 2003, \apj, 593, 772
\bibitem[Grindlay(1993)]{grind93}Grindlay, J.E. 1993, in The Globular Cluster-Galaxy Connection, ASP Conf. Ser. 48, 156
\bibitem[Grindlay et al.(2001)]{grind01}Grindlay, J.E., Heinke, C., Edmonds, P.D., \& Murray, S.S. 2001, Science, 292, 2290
\bibitem[Harris(1996)]{harris96}Harris, W.E. 1996, \aj, 112, 1487
\bibitem[Heinke et al.(2003)]{hein03}Heinke, C.O., Grindlay, J.E., Lugger, P.M., Cohn, H.N., Edmonds, P.D., Lloyd, D.A., \& Cool, A.M. 2003, \apj, 598, 501
\bibitem[Hut et al.(1992)]{hut92}Hut, P., et al. 1992, \pasp, 104, 1992
\bibitem[Hut \& Bahcall(1983)]{hut83} Hut, P., \& Bahcall, J.~N.\ 1983, \apj, 268, 319 
\bibitem[Irwin, Athey, \& Bregman(2003)]{irwin03}Irwin, J.A., Athey, A.E., \& Bregman, J.N. 2003, \apj, 587, 356
\bibitem[Ivanova et al.(2005)]{ivan05}Ivanova, N., Rasio, F.A., Lombardi, J.C., Dooley, K.L, \& Proulx, Z.F. 2005, \apjl, 621, L109
\bibitem[Jordan et al.(2005)]{jord05} Jordan, A., et al.\ 
2005, ArXiv Astrophysics e-prints, arXiv:astro-ph/0508219 
\bibitem[Kundu, Maccarone, \& Zepf(2002)]{kund02}Kundu, A., Maccarone, T.J., \& Zepf, S.E. 2002, \apjl, 574, L5
\bibitem[McLaughlin(2000)]{mcla00}McLaughlin, D.E. 2000, \apj, 539, 618
\bibitem[Noyola \& Gebhardt(2003)]{noyo03} Noyola, E., \& 
Gebhardt, K.\ 2003, Revista Mexicana de Astronomia y Astrofisica Conference 
Series, 18, 78
\bibitem[Paltrinieri et al.(2001)]{palt01} Paltrinieri, B., 
Ferraro, F.~R., Paresce, F., \& De Marchi, G.\ 2001, \aj, 121, 3114 
\bibitem[Piotto et al.(2003)]{piot03}Piotto, G., De Angeli, F., King, I.R., Djorgovski, S.G., Bono, G., Cassisi, S., Meylan, G., Recio-Blanco, A., Rich, R.M., \& Davies, M.B. 2003, \apj, 588, 464
\bibitem[Pooley et al.(2003)]{pool03}Pooley, D., et al. 2003, \apjl 591, L131
\bibitem[Richer et al.(2004)]{rich04}Richer, H.B., et al. 2004, \aj, 127, 2771
\bibitem[Rubenstein \& Bailyn(1997)]{rube97}Rubenstein, E.P., \& Bailyn, C.D. 1997, \apj, 474, 701
\bibitem[Salaris and Weiss(2002)]{sala02}Salaris, M., and Weiss, A. 2002, \aap, 388, 492
\bibitem[Verbunt(2001)]{verb01}Verbunt, F. 2001, \aap, 368, 137
\bibitem[Voges et al.(2000)]{voge00}Voges, W., Aschenbach, B., Boller, Th., Brauninger, H., Briel, U., Burkert, W., Dennerl, K., Englhauser, J., Gruber, R., Haberl, F., Hartner, G., Hasinger, G., Pfeffermann, E., Pietsch, W., Predehl, P., Schmitt, J., Trumper, J., Zimmermann, U. 2000, VizieR Cat. 9029
\end{thebibliography}
\end{document}